\documentstyle[epsf]{aipproc}
\newcommand{\be}{\begin{equation}}
\newcommand{\ee}{\end{equation}}  

\newcommand{\ber}{\begin{eqnarray}}
\newcommand{\eer}{\end{eqnarray}}
\begin{document}
\title { Large CP Violation in   
$B\to K^{(*)} X$ Decays}  
\author{\rm {T. E. Browder$^1$, A. Datta$^2$, X.-G. He$^3$ and 
S. Pakvasa$^1$} \\} 
\address {
 $^1$Department of Physics and Astronomy 
  University of Hawaii, 
 Honolulu, Hawaii 96822\\
 $^2$Department of Physics and Astronomy 
  University of Toronto, 
 Toronto, Ontario M5S 1A7, Canada\\
 $^3$ School of Physics
  University of Melbourne,
 Parkville, Victoria, 3052 
 Australia.
}
\maketitle 
\begin{abstract}
We consider the possibility of observing CP violation in quasi-inclusive
decays of the type $B^-\rightarrow K^- X$, $B^-\rightarrow K^{*-} X$,
 $\bar B^0\rightarrow K^- X$ and 
$\bar B^0\rightarrow K^{*-} X$, where $X$ does not contain
strange quarks. We present estimates of rates
and asymmetries for these decays in the Standard
Model and comment on the experimental feasibility of observing CP
violation in these decays at future $B$ factories. We find the rate
asymmetries can be quite sizeable.
\end {abstract}
\section{Introduction}
The possibility of observing large CP violating asymmetries in the
decay of $B$ mesons motivates the construction of high luminosity $B$
factories at several of the world's high energy physics laboratories.
The theoretical and the experimental signatures of these asymmetries
have been extensively 
discussed elsewhere\cite{BABAR},\cite{BELLE},\cite{Br},
\cite{Buras},\cite{dho}. 
At asymmetric $B$ factories, it is possible to measure the
time dependence of 
$B$ decays and therefore time dependent rate asymmetries of
neutral $B$ decays due to $B-\bar B$ mixing. The measurement of time 
dependent asymmetries in the exclusive modes $\bar{B}^0\to \psi K_s$
and $\bar{B}^0\to \pi^+\pi^-$ will allow the determination of the
angles in the Cabbibo-Kobayashi-Maskawa unitarity triangle. 
This type of CP violation
has been studied extensively in the literature.

Another type of CP violation also
exists in $B$ decays, direct CP violation in the 
$B$ decay amplitudes. This type of CP violation in $B$ decays has 
also been discussed by several authors although not as extensively.
For charged $B$ decays 
calculation of the magnitudes of the effects for some
exclusive modes and 
inclusive modes have been carried out\cite{Soni},\cite{HouW},
\cite{Wolfie},\cite{FSHe},
\cite{Du},\cite{Kamal},\cite{Kramer}. In contrast
to asymmetries induced by $B-{\overline B}$ mixing, the magnitudes have
large hadronic uncertainties, especially for the exclusive modes.
Observation of these asymmetries can be used to rule out the superweak
class of models\cite{superweak}.

In this paper we describe several quasi-inclusive experimental
signatures which could provide useful information on direct CP violation
at the high luminosity facilities of the future. One of the goals
is to increase the number of events available at experiments for
observing a CP asymmetry. In particular we examine the inclusive
decay of the neutral and the charged $B$ to either a
charged $K$ or a charged $K^*$ meson. By applying the
appropriate cut on the kaon (or $K^*$) energy one can isolate a signal 
with little background from $b\rightarrow c$ transitions.
Furthermore, these quasi-inclusive modes are expected to have less
hadronic uncertainty than
the exclusive modes, would have larger branching ratios and,
compared to the purely inclusive modes they may have larger CP
asymmetries. In this paper 
we will consider modes of the type $B \rightarrow
 K(K^*) X$ that have the strange quark only in the $K(K^*)$-meson.

In the sections which follow, we describe
the experimental signature and method. We then calculate
the rates and asymmetries for inclusive $B^-\to K^-(K^{*-})$ and 
$\bar{B}^0\to K^-(K^{*-})$
decays.

\section{Experimental Signatures for Quasi-Inclusive $b\to s g^*$}

In the $\Upsilon(4S)$ center of mass frame, 
the momentum of the $K^{(*)-}$ from quasi-two body $B$ decays
such as $B\to K^{(*)-} X$ may have momenta 
above the kinematic limit for $K^{(*)-}$ mesons from 
$b\to c$ transitions.
This provides an experimental signature for $b\to s g^*$, $g^*\to u \bar{u}$ 
or $g^*\to d \bar{d}$ decays where $g^*$ denotes a gluon. 
This kinematic separation between $b\to c$ 
and $b\to s g^*$ transitions is illustrated by
a generator level Monte Carlo simulation in Figure 1 for the case
of $B\to K^{*-}$. (The $B\to K^-$ spectrum will be similiar).
This experimental signature can be applied to the asymmetric energy
$B$ factories if one boosts backwards along the z axis into the 
$\Upsilon(4S)$ center of mass frame.

Since there is a large background (``continuum'')
from the non-resonant processes $e^+ e^-\to q \bar{q}$ where
$q=u, d, s, c$, experimental cuts on the event shape are also imposed. 
To provide additional continuum suppression, the  
``B reconstruction'' technique has been employed. The requirement
that the kaon and $n$ other pions form a system consistent in beam
constrained mass and energy with a 
$B$ meson dramatically reduces the background. After these
requirements are imposed, one searches for an excess in the kaon
momentum spectrum above the $b\to c$ region. 
Only one combination per event is chosen.
No effort is made to unfold the feed-across
between submodes with different values of n.

\begin{figure}[htb]
\centerline{\epsfysize 3.4 truein \epsfbox{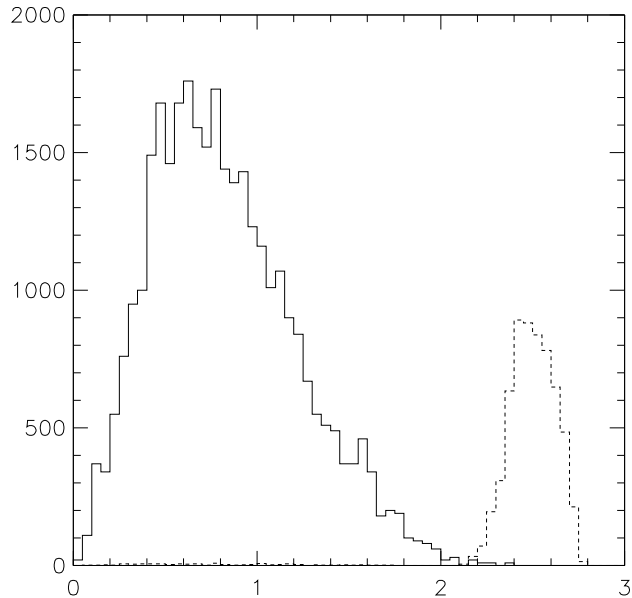}}
\caption{Generated Inclusive $B\to K^{*-}$ momentum spectrum.
The component below $2.0$ GeV/c is due to $b\to c$ decays while the
component above 2.0 GeV/c arises from quasi-two body $b\to s g^*$ decay.
The normalization of the $b\to c$ component is reduced by a factor of
approximately 100 so that both components are visible.}
\label{bkx_inclusive}
\end{figure}

Methods similar to these have been successfully used 
by the CLEO~II experiment to isolate a signal
in the inclusive single photon energy spectrum
and measure the branching fraction for inclusive $b\to s\gamma$ transitions
and to set upper 
limits on $b\to s \phi$ transitions\cite{cleobsg},\cite{cleophix}. 
It is clear from these studies that the $B$ reconstruction method provides
adequate continuum background suppression.

The decay modes that will be used here are listed below:
\begin{enumerate}
\item $B^-\to K^{(*)-}\pi^0$
\item $\bar{B}^0\to K^{(*)-} \pi^+$ 
\item $B^- \to K^{(*)-} \pi^-\pi^+$ 
\item $\bar{B}^0\to K^{(*)-}\pi^+\pi^0$
\item $\bar{B}^0\to K^{(*)-}\pi^+\pi^-\pi^+$
\item $B^-\to K^{(*)-} \pi^+\pi^-\pi^0$
\item $B^-\to K^{(*)-}\pi^+\pi^-\pi^+\pi^-$
\item $\bar{B}^0\to K^{(*)-}\pi^+\pi^-\pi^+\pi^0$
\end{enumerate}

In case of multiple entries for a decay mode, we choose the
best entry on the basis of a $\chi^2$ formed from the beam
constrained mass and energy difference
(i.e. $\chi^2= (M_B/\delta M_B)^2+ (\Delta E/\delta\Delta E)^2$). 
In case of multiple
decay modes per event, the best decay mode candidate is picked
on the basis of the same $\chi^2$.

Cross-feed between different $b\to s g$ decay modes 
(i.e. the misclassification of decay modes)  provided
the $K^{(*)-}$ is correctly identified, is not a concern 
as the goal is to extract an inclusive signal. The purpose of the
$B$ reconstruction method is to reduce continuum background.
As the multiplicity of the decay mode increases, however,
the probability of misrecontruction will increase.

The signal is isolated as excess $K^{(*)-}$ production in the high
momentum signal region ($2.0<p_{K^{(*)}}<2.7$ GeV) above continuum background.
To reduce contamination from high momentum 
$B\to \pi^- (\rho^-)$ production and
residual $b\to c$ background, we assume the presence of a high
momentum particle identification system as will be employed in
the BABAR, BELLE, and CLEO~III experiments.

We propose to measure the asymmetry
$N(K^{(*)+} - K^{(*)-})/N(K^{(*)+} + K^{(*)-})$ where $K^{(*)\pm}$ originates
from a partially reconstructed $B$ decay such as $B\to K^{(*)-} (n\pi)^0$
where the additional pions have net charge $0$ and $n\le 4$ and one neutral
pion is allowed and $2.7>p(K^{(*)-})>2.0$ GeV. We assume that the 
contribution from $B\to K^-\eta^{'} X $ 
decays has been removed by cutting on the $\eta^{'}$ region in
$X$ mass. It is possible that the anomalously large rate from
this source\cite{cleo_etaprime,theory_etaprime} could dilute the asymmetry.

\begin{figure}[htb]
\centerline{\epsfysize 5.0 truein \epsfbox{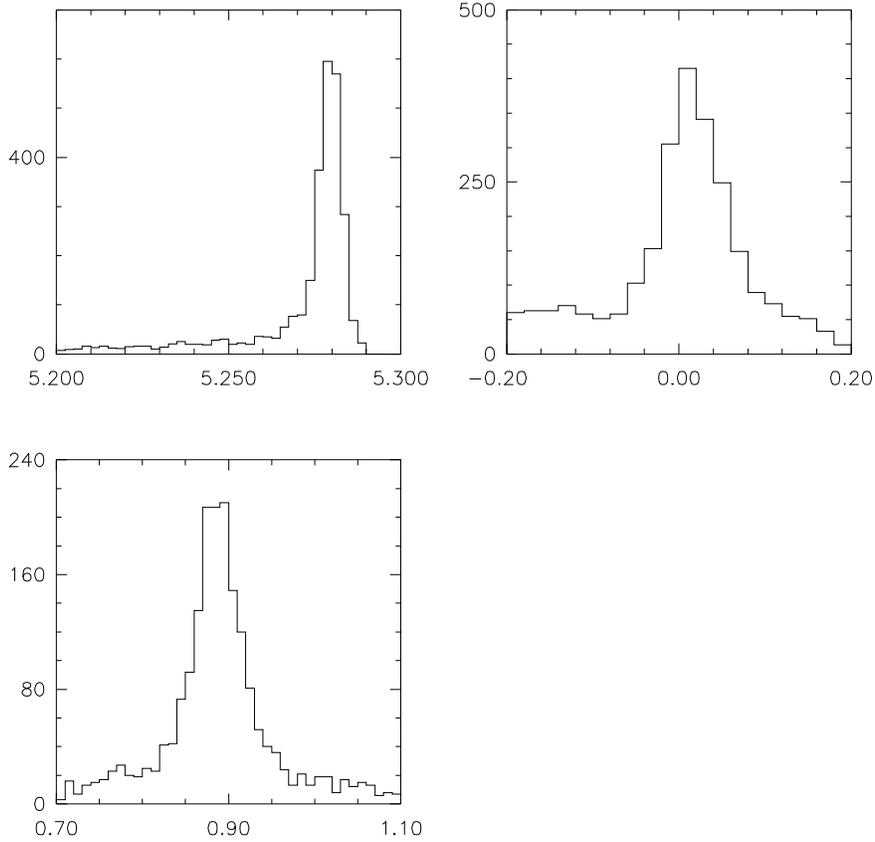}}
\caption{Monte Carlo simulation of inclusive $B\to K^{*-} X$ signal
with the B reconstruction method: (a) The beam constrained mass
distribution (b) The distribution of energy difference (c) The $K^-
\pi^0$ invariant mass after selecting on energy difference and beam
constrained mass}
\label{bkx_brecon}
\end{figure}

\section{ Effective Hamiltonian} 
In the Standard Model (SM) 
the amplitudes for hadronic $B$ decays of the type $b\to q \bar{f} f$ 
are generated by the following effective 
Hamiltonian \cite{Reina}:
\begin{eqnarray}
H_{eff}^q &=& {G_F \over \protect \sqrt{2}} 
   [V_{fb}V^*_{fq}(c_1O_{1f}^q + c_2 O_{2f}^q) -
     \sum_{i=3}^{10}(V_{ub}V^*_{uq} c_i^u
+V_{cb}V^*_{cq} c_i^c +V_{tb}V^*_{tq} c_i^t) O_i^q] +H.C.\;,
\end{eqnarray}
where the
superscript $u,\;c,\;t$ indicates the internal quark, $f$ can be $u$ or 
$c$ quark. $q$ can be either a $d$ or a $s$ quark depending on 
whether the decay is a $\Delta S = 0$
or $\Delta S = -1$ process.
The operators $O_i^q$ are defined as
\begin{eqnarray}
O_{f1}^q &=& \bar q_\alpha \gamma_\mu Lf_\beta\bar
f_\beta\gamma^\mu Lb_\alpha\;,\;\;\;\;\;\;O_{2f}^q =\bar q
\gamma_\mu L f\bar
f\gamma^\mu L b\;,\nonumber\\
O_{3,5}^q &=&\bar q \gamma_\mu L b
\bar q' \gamma_\mu L(R) q'\;,\;\;\;\;\;\;\;O_{4,6}^q = \bar q_\alpha
\gamma_\mu Lb_\beta
\bar q'_\beta \gamma_\mu L(R) q'_\alpha\;,\\
O_{7,9}^q &=& {3\over 2}\bar q \gamma_\mu L b  e_{q'}\bar q'
\gamma^\mu R(L)q'\;,\;O_{8,10}^q = {3\over 2}\bar q_\alpha
\gamma_\mu L b_\beta
e_{q'}\bar q'_\beta \gamma_\mu R(L) q'_\alpha\;,\nonumber
\end{eqnarray}
where $R(L) = 1 \pm \gamma_5$, 
and $q'$ is summed over u, d, and s.  $O_1$ are the tree
level and QCD corrected operators. $O_{3-6}$ are the strong gluon induced
penguin operators, and operators 
$O_{7-10}$ are due to $\gamma$ and Z exchange (electroweak penguins),
and ``box'' diagrams at loop level. The Wilson coefficients
 $c_i^f$ are defined at the scale $\mu \approx m_b$ 
and have been evaluated to next-to-leading order in QCD.
The $c^t_i$ are the regularization scheme 
independent values obtained in Ref. \cite{FSHe}.
We give the non-zero  $c_i^f$ 
below for $m_t = 176$ GeV, $\alpha_s(m_Z) = 0.117$,
and $\mu = m_b = 5$ GeV,
\begin{eqnarray}
c_1 &=& -0.307\;,\;\; c_2 = 1.147\;,\;\;
c^t_3 =0.017\;,\;\; c^t_4 =-0.037\;,\;\;
c^t_5 =0.010\;,
 c^t_6 =-0.045\;,\nonumber\\
c^t_7 &=&-1.24\times 10^{-5}\;,\;\; c_8^t = 3.77\times 10^{-4}\;,\;\;
c_9^t =-0.010\;,\;\; c_{10}^t =2.06\times 10^{-3}\;, \nonumber\\
c_{3,5}^{u,c} &=& -c_{4,6}^{u,c}/N_c = P^{u,c}_s/N_c\;,\;\;
c_{7,9}^{u,c} = P^{u,c}_e\;,\;\; c_{8,10}^{u,c} = 0
\end{eqnarray}
where $N_c$ is the number of color. 
The leading contributions to $P^i_{s,e}$ are given by:
 $P^i_s = ({\frac{\alpha_s}{8\pi}}) c_2 ({\frac{10}{9}} +G(m_i,\mu,q^2))$ and
$P^i_e = ({\frac{\alpha_{em}}{9\pi}})
(N_c c_1+ c_2) ({\frac{10}{9}} + G(m_i,\mu,q^2))$.  
The function
$G(m,\mu,q^2)$ is given by
\begin{eqnarray}
G(m,\mu,q^2) = 4\int^1_0 x(1-x)  \mbox{ln}{m^2-x(1-x)q^2\over
\mu^2} ~\mbox{d}x \;.
\end{eqnarray}
All the above coefficients are obtained up to one loop order in electroweak 
interactions. The momentum $q$ is the momentum carried by the virtual gluon in
the penguin diagram.
When $q^2 > 4m^2$, $G(m,\mu,q^2)$ becomes imaginary. 
In our calculation, we 
use $m_u = 5$ MeV, $m_d = 7$ MeV, $m_s = 200$ MeV, $m_c = 1.35$ GeV
\cite{lg,PRD}.

We assume that the final state phases calculated at the quark level
will be a good approximation
 to the sizes and the signs of the FSI phases at
the hadronic level for quasi-inclusive decays
when the final state particles are quite energetic
as is the case for the $B$ decays in the kinematic range
of experimental interest\cite {Soni}.

\section{Matrix Elements for $ B^-  \rightarrow {K^-}  X $
and $\bar{B}^0  \rightarrow {K^-}  X $}

We proceed to calculate 
 the matrix elements of the form $<K X|H_{eff}|B>$ which represents 
the process  $ B\rightarrow K X$ and where $H_{eff}$ has been described 
above. The effective Hamiltonian consists of operators with a
 current $\times$ current structure. 
Pairs of such operators can be expressed in terms of
color singlet and color octet structures which lead to color singlet and
color octet matrix elements. In the factorization approximation,
one separates out the currents in the operators by inserting the vacuum
state and neglecting any QCD interactions between the two currents. The
basis for this approximation is that, if the quark pair created by one of
the currents carries large energy then it will not have significant 
QCD interactions. In this approximation the color octet matrix element
does not contribute because it cannot be expressed in a factorizable
color singlet form. 
In our case, since the
energy of the quark pairs that either creates the $K$ or the $X$ state
is rather large, factorization is likely to be a good first 
approximation. To accommodate some 
deviation from this approximation we treat
$N_c$, the number of colors that enter in the calculation of the matrix
elements, as a free parameter. In our calculation we will see how our
results vary with different choices of $N_c$. The value of $N_c\sim 2$ is
suggested by experimental data on 
low multiplicity hadronic $B$ decays\cite{Br}. The amplitude
for $B \to K^{(*)} X$ can in general be split into
a three body and a two body part.
Detailed expressions for the matrix elements, decay distributions and asymmetries can be found in \cite{Qincl}

\section{Results and Discussion }
 In this section we discuss the results of our calculations. 
We find that there can be significant asymmetries in $B \to K (K^*) X$
decays especially in the region  $E_K> 2 $ GeV which is also the region
where an experimental  signal for such decays can be isolated.
The branching ratios are of order $O(10^{-4})$ 
which are within reach for future B factories.
The contribution of the amplitude with the top quark in the loop
accounts for 60-75\% of the inclusive branching fraction. However,
since the top quark amplitude is large and has no absorptive part
in contrast to the c quark amplitude, 
the top quark contribution reduces the net CP asymmetry from 30-50\% to
about 10\%. This calculation includes the contribution from
electroweak penguins. We find that the electroweak penguin 
contributions increase the decay rates by 10-20\% but reduce
the overall asymmetry by 20-30\%. The main sources of uncertainties in our
calculation are discussed extensively
in \cite{Qincl}.

\begin{figure}[htb]
\centerline{\epsfysize 2.8 truein \epsfbox{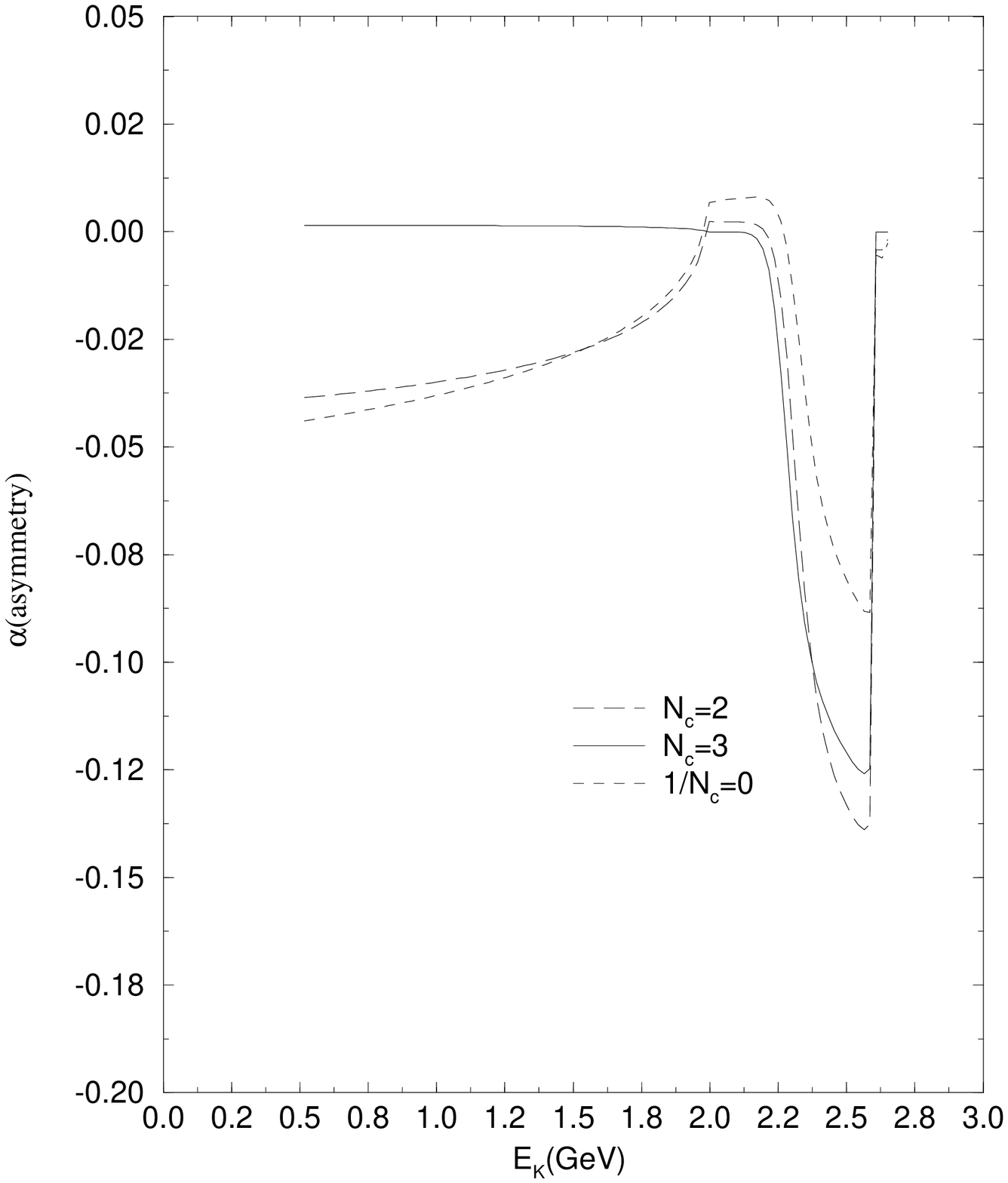} ~~~~
\epsfysize 2.8 truein \epsfbox{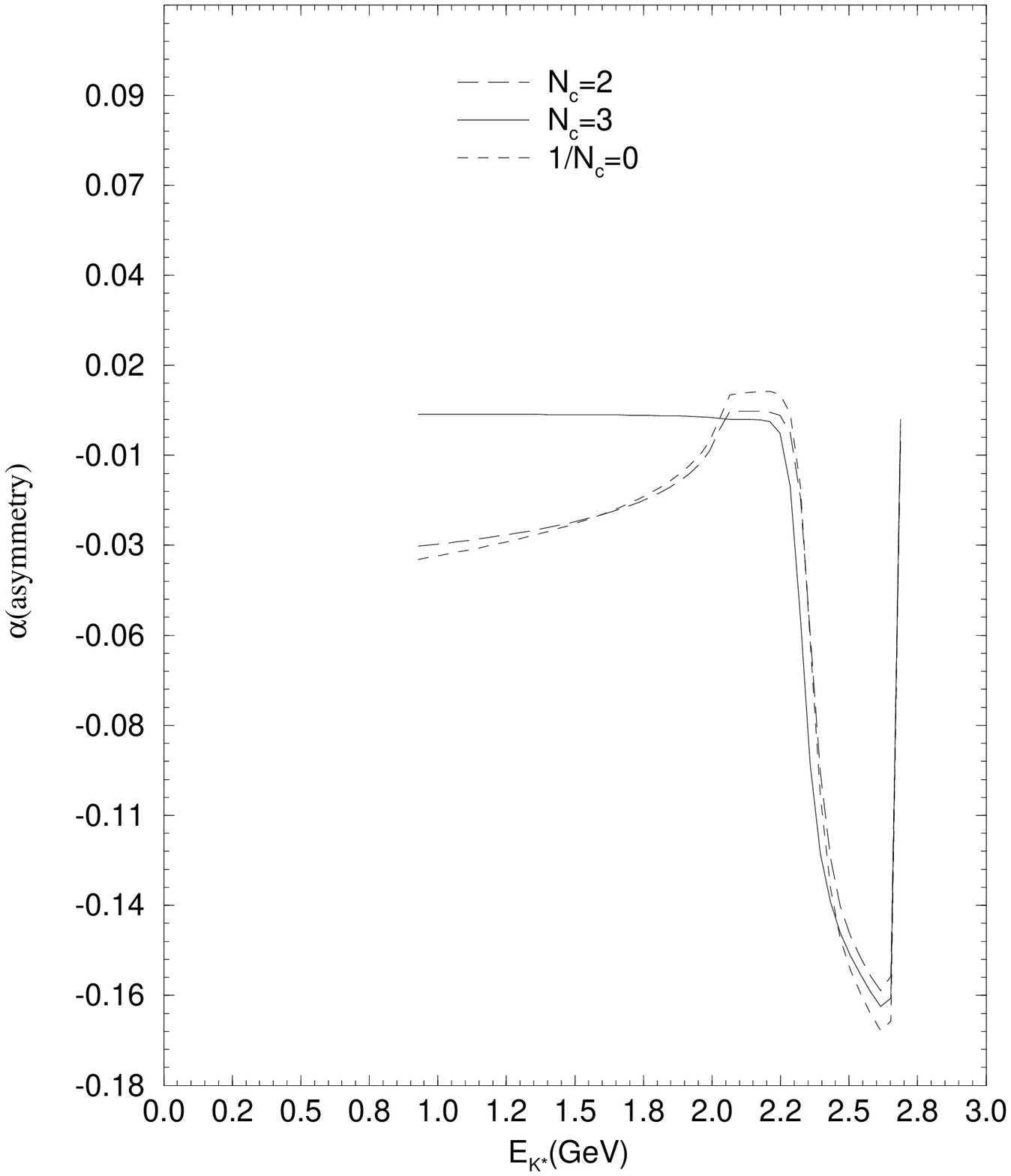}}
\caption{Predicted Asymmetries for 
$B^-\to K^- X$ and $B^-\to K^{*-} X$ as a function of the
kaon energy. The three sets of curves
indicate the sensitivity of the asymmetry to the value of $N_c$.
The values $N_c=2, 3, \infty$ are considered.}
\label{bkxasym_Nc}
\end{figure}

 The asymmetries are sensitive to the values of 
the Wolfenstein parameters $\rho$ and $\eta $. The 
existing constraints on the  values of $\rho$ and
$\eta$ come from measurements of $|V_{ub}|/|V_{cb}|$, $\epsilon_K$ in the
K system and $\Delta M_{B_d}$. (See  Ref.~\cite{Ali1} for a recent review).
In our calculation we will use $f_B =170$ MeV  and choose
$(\rho=-0.15,\eta=0.33)$. 

In Fig. 3 we
 show the asymmetries for $K$ and $K^*$ in the final
state in charged $B$ decays for different values of $N_c$.
Variation of the asymmetries with the different inputs in our calculation are 
presented in detail in \cite{Qincl}.
 
In Table.~\ref{Tb_integrated}
we give the branching fractions and the integrated
asymmetries for the inclusive decays for different $N_c$, 
$q^2=m_b^2/2$ ( $q$ is the gluon momentum
in the two body part of the amplitude ),
$f_B=170$ MeV, $\rho=-0.15, \eta=0.33 $. For the charged $B$ decays we
also show the decay rates and asymmetries for 
$E_K>2 $ ($2.1$) GeV as that is the region of the signal.
%

%
\begin{table}
\caption{Integrated decay rates and asymmetries for $B\to K^{(*)} X$ Decay}
\begin{center}
\begin{tabular}{ccc}
 Process &  Branching Ratio ($1.65\times 10^{-4}$) &  
Integrated Asymmetry  \\ \hline
    &  & \\
$B^-\rightarrow K^- X$ & $1.02,0.79,1.20$ &$-0.10,-0.11,-0.050$\\

$B^-\rightarrow K^- X (E_K \ge 2.1 \rm{GeV})$ & $0.81,0.74,0.77$
&$-0.12,-0.12,-0.07$\\

${\overline B^0}\rightarrow K^- X$ 
 & $0.6,0.7,0.8$ & $-0.12,-0.12,-0.13$\\
%
     &  & \\
$B^-\rightarrow K^{*-} X$ & $1.37,1.24,2.30$ & $-0.11,-0.14,-0.11$\\

$B^-\rightarrow K^{*-} X (E_{K^*} \ge 2.1 \rm{GeV})$
 & $1.05,1.16,1.67$ &$-0.14,-0.15,-0.14$\\

${\overline B^0}\rightarrow K^{*-}X$  
& $1.05,1.16,1.39$ & $-0.15,-0.15,-0.16$\\
\end{tabular}
\end{center}
\label{Tb_integrated}
\end{table}

The above figures
show that there can be significant asymmetries in $B \to K^{(*)} X$
decays, especially in the region  $E_K> 2 $ GeV which is the region
of experimental sensitivity 
 for such decays.
 As already mentioned, our calculation is
not free of theoretical uncertainties. Two strong assumptions used in
our calculation are the use of quark level strong phases for the FSI
phases at the hadronic level and the choice of the
 value of the gluon momentum $q^2$ in the two body decays. Other
uncertainties from the use of different heavy to light
form factors, the use of factorization, the model of the
B meson wavefunction, the value of the charm quark
mass and the choice of the renormalization scale $\mu$ 
have smaller effects on the asymmetries \cite{Qincl}.

\section{Conclusion}

We find significant direct CP violation in the inclusive decay
$B\to K^{-} X$ and $B\to K^{*-} X$ for $2.7>E_{K^{(*)}}>2.0$ GeV.
The branching fractions are in the $10^{-4}$ range and the CP
asymmetries may be sizeable. These asymmetries should be observable
at future $B$ factories and could be used to rule out the superweak
class of models.

\section{Acknowledgements}

This work was supported in part by National Science and Engineering Research Council of 
Canada (A. Datta).
A. Datta thanks the organisers of M.R.S.T for hospitality 
and an interesting conference.

\end{document}